\documentclass[conference]{IEEEtran}
\usepackage{ifpdf}
\ifpdf
    \usepackage[pdftex]{graphicx}
    \graphicspath{{figs/}{matlab/}}   % still needed, even with Inkscape-Pdf-Latex
    \usepackage[update]{epstopdf}
\else
	\usepackage{graphicx}
	\graphicspath{{figs/}{matlab/}}   % still needed, even with Inkscape-Pdf-Latex
\fi

\usepackage{array}
\usepackage{amsmath}
\usepackage{amsfonts}
\usepackage{amssymb}
\usepackage{amstext}
\usepackage{latexsym}
\usepackage{color}
\usepackage{cite,url}
\usepackage{setspace}
\newtheorem{thm}{Theorem}
\newtheorem{prop}{Proposition}
\newtheorem{lemma}{Lemma}

\newtheorem{coro}[lemma]{Corollary}

\usepackage{pdflscape}
\allowdisplaybreaks
\usepackage{multirow}

\DeclareMathSizes{10}{9}{8}{7} 

\begin{document}

\title{Multilevel Diversity Coding with Regeneration: Separate Coding Achieves the MBR Point}
\author{\IEEEauthorblockN{Shuo Shao and Tie Liu}
\IEEEauthorblockA{Dept. of Electrical \& Computer Engineering\\
Texas A\&M University, College Station TX\\
Email: \{shaoshuo,tieliu\}@tamu.edu}
\and
\IEEEauthorblockN{Chao Tian}
\IEEEauthorblockA{Dept. of Electrical Engineering \& Computer Science\\
University of Tennessee, Knoxville TN\\
Email: chao.tian@utk.edu}
}

\textfloatsep=0.15cm
\intextsep=0.15cm
\abovecaptionskip=0.15cm
\belowcaptionskip=0.15cm
\setlength{\abovedisplayskip}{5pt}
\setlength{\belowdisplayskip}{5pt}

\maketitle
\begin{abstract}

The problem of multilevel diversity coding with regeneration is considered in this work. Two new outer bounds on the optimal tradeoffs between the normalized storage capacity and repair bandwidth are established, by which the optimality of separate coding at the minimum-bandwidth-regeneration (MBR) point follows immediately. This resolves a question left open in a previous work by Tian and Liu.
\end{abstract}
%\vspace{-0.2cm}
\begin{IEEEkeywords}
Information theory, distributed storage, multilevel diversity coding, regenerating codes
\end{IEEEkeywords}


% \fi
%\IEEEpeerreviewmaketitle

%\vspace{-0.2cm}

\section{Introduction}
Diversity\let\thefootnote\relax\footnotetext{This work was supported in part by the National Science Foundation under Grants CCF-13-20237, CCF-15-24839, and CCF-15-26095.} 
 coding and node repair are two fundamental ingredients of reliable distributed storage systems. This paper considers the problem of $(n,d)$ multilevel diversity coding with regeneration (MLDR), which was first introduced in \cite{TL-IT15}. In this problem, a total of $d$ independent messages $M_1,\ldots,M_d$ of $B_1,\ldots,B_d$ bits, respectively, are to be stored in $n>d$ nodes each of capacity $\alpha$ bits. Two requirements need to be satisfied: (i) \textbf{Diversity reconstruction}: For any $k=1,\ldots,d$, the message $M_k$ can be recovered by accessing any $k$ (out of the total $n$) storage nodes, and (ii) \textbf{Node regeneration}: For any $i=1,\ldots,n$, the data stored at node $i$ can be regenerated by extracting $\beta$ bits of information each from any $d$ other nodes. We call such a code an $(n,d,(B_1,\ldots,B_d),(\alpha,\beta))$ MLDR code. A normalized storage-capacity {\em vs}. repair-bandwidth pair $(\bar{\alpha},\bar{\beta})$ is said to be {\em achievable} for a given normalized message size tuple $(\bar{B}_1,\cdots,\bar{B}_d)$ if an $(n,d,(B_1,\ldots,B_d),(\alpha,\beta))$ MLDR code can be found such that 
$\bar{B}_k=\frac{B_k}{\sum_{j=1}^{d}B_j}$ for $k=1,2\ldots,d$, $\bar{\alpha}=\frac{\alpha}{\sum_{j=1}^{d}B_j}$ and $\bar{\beta}=\frac{\beta}{\sum_{j=1}^{d}B_j}$.
A precise mathematical description of the problem can be found in \cite{TL-IT15}.

A natural strategy for this problem is to encode each individual message separately using an exact-repair regenerating code \cite{Dimakis-IT10,Kumar-IT11} of necessary parameters. More precisely, suppose that each message $M_k$ is encoded using an $(n,k,d,B_k,(\alpha_k,\beta_k))$ exact-repair regenerating code ({\em i.e.}, when storing a single message $M_k$ of $B_k$ bits, accessing any $k$ nodes of $\alpha_k$ capacity each can reconstruct $M_k$, and any node can be regenerated by extracting $\beta_k$ bits data each from any $d$ other nodes). Then, we have $\alpha = \sum_{k=1}^d \alpha_k$ and $\beta=\sum_{k=1}^d \beta_k$ for the resulting MLDR code. Let us define the individually normalized storage-capacity {\em vs}. repair-bandwidth pair as $(\bar{\alpha}_k,\bar{\beta}_k):= (\frac{{\alpha}_k}{B_k},\frac{{\beta}_k}{B_k})$ for $k=1,\ldots,d$. The overall storage-capacity {\em vs}. repair-bandwidth pair achieved by the separate coding scheme for the MLDR problem is thus given by:  
\begin{align}
(\bar{\alpha},\bar{\beta})=\left(\sum_{k=1}^{d}\bar{\alpha}_k\bar{B}_k,\sum_{k=1}^{d}\bar{\beta}_k\bar{B}_k\right).\label{eq:SC}
\end{align}
A fundamental problem of interest is whether separate coding can achieve the {\em optimal} tradeoffs between the normalized storage capacity and repair bandwidth for the MLDR problem.

\begin{figure}[t!]
\centering
\includegraphics[width=0.8\linewidth,draft=false]{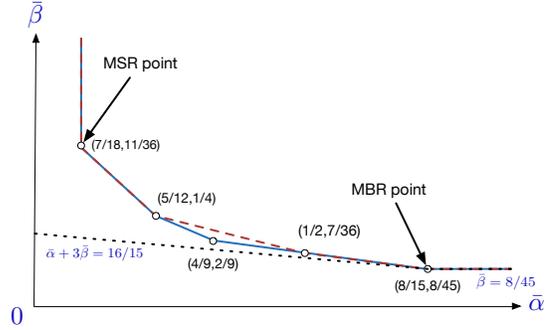}
\caption{The optimal tradeoff curve between the normalized storage capacity and repair bandwidth (the solid line) and the best possible tradeoffs that can be achieved by separate coding (dashed line) for the $(4,3)$ MLDR problem with $(\bar{B}_1,\bar{B}_2,\bar{B}_3)=(0,1/3,2/3)$ \cite{TL-IT15}. The two new outer bounds \eqref{eq:B1} and \eqref{eq:B2} intersect precisely at the MBR point.}
%\vspace{-0.3cm}
\label{fig}
\end{figure}

This question was first answered in \cite{TL-IT15}, where it was shown that separate coding is in general {\em suboptimal}. For concreteness, Figure~\ref{fig} shows the optimal tradeoff curve between the normalized storage capacity and repair bandwidth and the best possible tradeoffs that can be achieved by separate coding (see \cite{TL-IT15}) with $n=4$, $d=3$, and $(\bar{B}_1,\bar{B}_2,\bar{B}_3)=(0,1/3,2/3)$. As illustrated in Figure~\ref{fig}, separate coding is suboptimal when $\bar{\alpha}\in(5/12,1/2)$. On the other hand, when $\bar{\alpha} \leq 5/12$ or $\bar{\alpha} \geq 1/2$, separate coding can in fact achieve the optimal tradeoffs. In particular, for this example, separate encoding achieves the minimum-storage-regenerating (MSR) point $(7/18,11/36)$ and the minimum-bandwidth-regenerating (MBR) point $(8/15,8/45)$. In addition, it was shown in \cite{TL-IT15} that the optimality of separate coding at the MSR point is {\em not} a coincidence and in fact holds for {\em any} MLDR problem. It is thus natural to ask whether the same generalization holds for the MBR point as well; this problem was left open in \cite{TL-IT15}.

In this paper, we proved two new outer bounds on the optimal tradeoffs between the normalized storage-capacity and repair-bandwidth for general MLDR problem, by which the optimality of separate coding at the MBR point follows immediately. Our proofs are based on the classical ``peeling" argument, which sequentially removes the effects of certain coding requirements by grouping the corresponding random variables under the conditional terms. The technique was first introduced in \cite{RYH-IT97} and subsequently used in \cite{TL-IT15} to prove the optimality of separate coding at the MSR point. The telescoping results here, however, are much more involved than those proved in \cite{RYH-IT97} and \cite{TL-IT15}. 

{\em Notation}. For brevity, let $[i:j]:=\{i,i+1,\ldots,j\}$ for any positive integers $i\leq j$, $J_d :=\sum_{i=1}^{d}i$ for any integer $d\geq 0$, and $T_{d,k} :=\frac{1}{\sum_{i=1}^{k}(d+1-i)}$ for any integers $d \geq 1$ and $k\in [1:d]$. Without loss of generality, we assume $n \geq 2$ and $d\leq n-1$. 

\section{Main Results}
\begin{thm}\label{thm}
Any achievable normalized storage-capacity {\em vs}. repair-bandwidth pair $(\bar{\alpha},\bar{\beta})$ for the MLDR problem must satisfy:
\begin{align}
\bar{\beta} & \geq \sum_{k=1}^{d}T_{d,k}\bar{B}_k\label{eq:B1}\\
\mbox{and} \quad \bar{\alpha}+J_{d-1}\bar{\beta} & \geq J_{d}\sum_{k=1}^{d}T_{d,k}\bar{B}_k.\label{eq:B2}
\end{align}
\end{thm}

When set as equalities, the intersection of \eqref{eq:B1} and \eqref{eq:B2} is given by:
\begin{align*}
\left(\bar{\alpha},\bar{\beta}\right)%&=\left((J_d-J_{d-1})\sum_{k=1}^{d}T_{d,k}\bar{B}_k,\sum_{k=1}^{d-1}T_{d,k}\bar{B}_k\right)\\
&=\left(d\sum_{k=1}^{d}T_{d,k}\bar{B}_k,\sum_{k=1}^{d}T_{d,k}\bar{B}_k\right).
\end{align*}
For any $k\in [1:d]$, the MBR point for the $(n,k,d)$ exact-repair problem can be written as \cite{Dimakis-IT10}
\begin{align*}
\left(\bar{\alpha}_k,\bar{\beta}_k\right)=\left(dT_{d,k},T_{d,k}\right).
\end{align*}
By \eqref{eq:SC}, we immediately have the following corollary.
\begin{coro}
Separate coding achieves the MBR point for the MLDR problem.
\end{coro}

When $n=4$, $d=3$, and $(\bar{B}_1,\bar{B}_2,\bar{B}_3)=(0,1/3,2/3)$, the outer bounds \eqref{eq:B1} and \eqref{eq:B2} can be explicitly evaluated as $\bar{\beta} \geq 8/45$ and $\bar{\alpha}+3\bar{\beta}\geq16/15$, respectively. As illustrated in Figure~\ref{fig}, they intersect precisely at the MBR point $(8/15,8/45)$. Interestingly, for this example at least, the outer bound \eqref{eq:B2} is tight {\em only} at the MBR point.

\subsection{Proof of Theorem~\ref{thm} via Peeling Arguments}
\begin{figure}[t!]
\centering
\includegraphics[width=0.65\linewidth,draft=false]{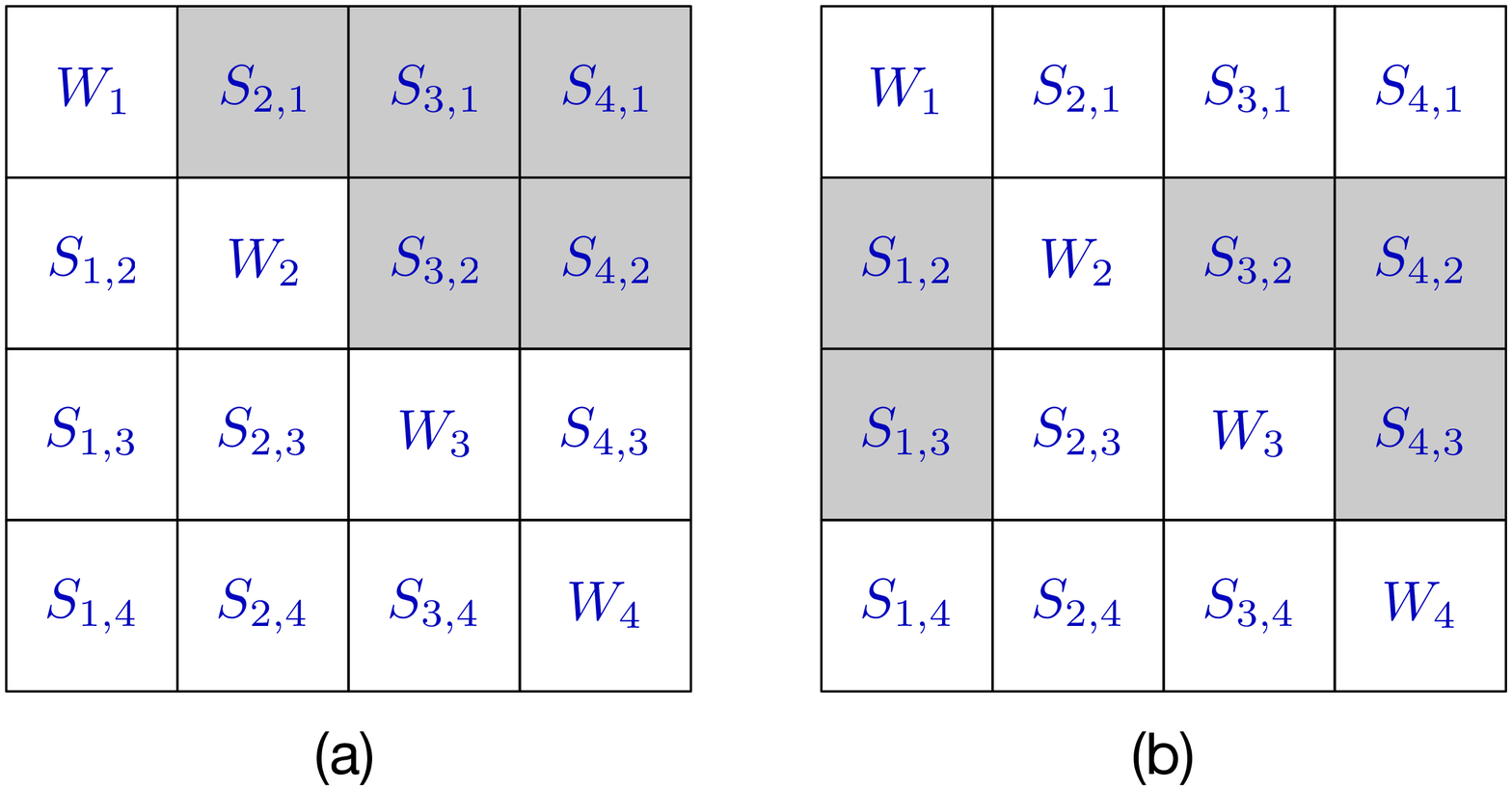}
\caption{The repair diagram of Duursma \cite{Duursma-P14} for $n=4$ and $d=3$. The key data structures (a) $l^{(2)}$ and (b) $l'_{[2:3]}$ are illustrated as the collections of shaded variables.}
%\vspace{-0.3cm}
\label{fig2}
\end{figure}

To prove the outer bounds \eqref{eq:B1} and \eqref{eq:B2}, we may fix $d\geq 1$ and assume, without loss of generality, that $n=d+1$. This is because if $n>d+1$, then the subsystem consisting of the first $d+1$ storage nodes forms an $(n'=d+1,d)$ MLDR problem, which also needs to satisfy the same set of constraints. 

The data stored at node $k$, $k=1,\ldots,n$, are denoted as $W_k$; the set $\{W_1,W_2,\ldots,W_k\}$ is written as $W^{(k)}$. The data extracted from node $j$ to regenerate node-$k$ is denoted as $S_{j,k}$. Let $S_{\tau,k}:=\{S_{j,k}:j\in \tau\}$ for any $k\in[1:d+1]$ and any $\emptyset \neq \tau\subseteq [1:d+1]-\{k\}$. Furthermore, let $l_0:=\emptyset$, $l_k:=S_{[k+1:d+1],k}$ for any $k=[1:d]$, and $l_{[i:j]}:=\{l_k:k\in[i,j]\}$ for any $1\leq i \leq j \leq d$. As we shall see, the data structures $l^{(k)}:=l_{[1:k]}$ for $k\in[1:d]$ and $\{W_1,l_{[2:k]}\}$ for $k\in [2:d]$, which are closely related to the repair diagram introduced by Duursma \cite{Duursma-P14}, play a key role in the peeling arguments for proving the outer bounds \eqref{eq:B1} and \eqref{eq:B2}; see Figure~\ref{fig2}.

Due to the built-in symmetry in the problem, we only need to consider the so-called {\em symmetrical codes} \cite{Tian-JSAC13} when discussing the optimal tradeoffs between the normalized storage capacity and repair bandwidth for the MLDR problem. For symmetrical codes, the joint entropy of any subset of random variables from $W^{(d+1)}\bigcup \{S_{j,k}: j,k\in[1:d+1],j\neq k\}\bigcup M^{(d)}$ remains {\em unchanged} under any permutation over the storage-node indices. Further note that $l_j$ is invariant (i.e., the collection of random variables from $l_j$ remains unchanged) under any permutation $\pi$ over $[1:d+1]$ such that $\pi(i)=i$ for $i\in[1:j]$; this fact is used repeatedly in proving the following telescoping results.

\begin{prop}[Telescoping over $l^{(k)}$]\label{prop:P1}
For any symmetrical $(n,d,(B_1,\ldots,B_d),(\alpha,\beta))$ MLDR code with $n=d+1$ and any $k\in[1:d-1]$, we have
\begin{align}
T_{d,k}H(l^{(k)}|M^{(k)}) \geq T_{d,k+1}H(l^{(k+1)}|M^{(k)}).\label{eq:P1}
\end{align}
\end{prop}

\begin{prop}[Telescoping over $\{W_1,l_{[2:k]}\}$]\label{prop:P2}
For any symmetrical $(n,d,(B_1,\ldots,B_d),(\alpha,\beta))$ MLDR code with $n=d+1$ and any $k\in[1:d-1]$, we have
\begin{align}
H&(W_1,l_{[2:k]}|M^{(k)})+\nonumber\\
&(d-k)T_{n,k}H(l^{(k)}|M^{(k)})\geq H(W_1,l_{[2:k+1]}|M^{(k)}).\label{eq:P2}
\end{align}
\end{prop}

With the help of the above telescoping results, we can now prove Theorem~\ref{thm} using the peeling arguments as follows.

\begin{IEEEproof}[Proof of \eqref{eq:B1}] 
%Let us first show that for any symmetrical $(n,d,(B_1,\ldots,B_d),(\alpha,\beta))$ MLDR code with $n=d+1$ and any $k\in[1:d]$, we have
We shall prove the following bound
\begin{align}
\beta \geq \sum_{j=1}^{k}T_{d,j}B_j+T_{d,k}H(l^{(k)}|M^{(k)})\label{eq:Induc1}
\end{align}
by induction. Note that
\begin{align*}
\beta & \stackrel{(a)}{\geq} \frac{1}{d}\sum_{i=2}^{d+1}H(S_{i,1}) \stackrel{(b)}{\geq} T_{d,1}H(l_1)\\
& \stackrel{(c)}{=} T_{d,1}H(l_1,M_1) \stackrel{(d)}{=} T_{d,1}H(M_1)+T_{d,1}H(l_1|M_1),
\end{align*}
and thus \eqref{eq:Induc1} holds for $k=1$. Here, $(a)$ follows from the repair-bandwidth constraints $H(S_{i,1}) \leq \beta$ for $i\in[2:d+1]$ and the fact that $T_{d,1}=1/d$; $(b)$ is due to the union bound on entropy; $(c)$ follows from the fact that $M_1$ is a function of $W_1$, thus a function of $l_1$; and $(d)$ is due to the chain rule for entropy.

Now assume that  \eqref{eq:Induc1} holds for some $k\in[1:d-1]$. Substituting the telescoping result \eqref{eq:P1} into \eqref{eq:Induc1}, we have
\begin{align*}
\beta &\geq \sum_{j=1}^{k}T_{d,j}B_j+T_{d,k+1}H(l^{(k+1)}|M^{(k)})\\
&\stackrel{(a)}{=} \sum_{j=1}^{k}T_{d,j}B_j+T_{d,k+1}H(l^{(k+1)},M_{k+1}|M^{(k)})\\
&\stackrel{(b)}{=} \sum_{j=1}^{k}T_{d,j}B_j+T_{d,k+1}H(M_{k+1}|M^{(k)})+\\
& \hspace{13pt} T_{d,k+1}H(l^{(k+1)}|M^{(k)},M_{k+1})\\
&\stackrel{(c)}{=} \sum_{j=1}^{k+1}T_{d,j}B_j+T_{d,k+1}H(l^{(k+1)}|M^{(k+1)}),
\end{align*}
which completes the induction and hence the proof of \eqref{eq:Induc1}. Here, $(a)$ follows from the fact that $M_{k+1}$ is a function of $W^{(k+1)}$, which is turn a function of $l^{(k+1)}$; $(b)$ is by  the chain rule for conditional entropy; and $(c)$ follows from the facts that all messages are independent and that $H(M_{k+1})=B_{k+1}$.

Setting $k=d$ in \eqref{eq:Induc1} and by the fact that $H(l^{(d)}|M^{(d)}) \geq 0$, we have
\begin{align}
\beta \geq \sum_{j=1}^{d}T_{d,j}B_j. \label{eq:Temp1}
\end{align}
Normalizing both sides of \eqref{eq:Temp1} by $\sum_{k=1}^{d}B_k$ completes the proof of the outer bound \eqref{eq:B1}. %\hfill\IEEEQED
\end{IEEEproof}

\begin{IEEEproof}[Proof of \eqref{eq:B2}]
We shall prove that 
\begin{align}
\alpha+J_{d-1}\beta &\geq J_{d}\sum_{j=1}^{k}T_{d,j}B_j+H(W_1,l_{[2:k]}|M^{(k)})+\nonumber\\
& \hspace{13pt} J_{d-k}T_{d,k}H(l^{(k)}|M^{(k)}),\label{eq:Induc2}
\end{align}
by induction. Note that
\begin{align*}
&\alpha+J_{d-1}\beta\\
& \stackrel{(a)}{\geq} H(W_1)+\frac{J_{d-1}}{d}\sum_{i=2}^{d+1}H(S_{i,1})\\
& \stackrel{(b)}{\geq} H(W_1)+J_{d-1}T_{d,1}H(l_1)\\
& \stackrel{(c)}{=} H(W_1,M_1)+J_{d-1}T_{d,1}H(l_1,M_1)\\
& \stackrel{(d)}{=} H(M_1)+H(W_1|M_1)+J_{d-1}T_{d,1}H(M_1)+\\
& \hspace{13pt} J_{d-1}T_{d,1}H(l_1|M_1)\\
& \stackrel{(e)}{=} (1+J_{d-1}T_{d,1})B_1+H(W_1|M_1)+J_{d-1}T_{d,1}H(l_1|M_1)\\
& \stackrel{(f)}{=} J_{d}T_{d,1}B_1+H(W_1|M_1)+J_{d-1}T_{d,1}H(l_1|M_1),
\end{align*}
and thus \eqref{eq:Induc2} holds for $k=1$. Here, $(a)$ follows from the storage-capacity constraint $H(W_1) \leq \alpha$ and the repair-bandwidth constraints $H(S_{i,1}) \leq \beta$ for $i\in[2:d+1]$; $(b)$ is by the union bound on entropy and the fact that $T_{d,1}=1/d$; $(c)$ follows from the fact that $M_1$ is a function of $W_1$, thus a function of $l_1$; $(d)$ is due to the chain rule for entropy; $(d)$ is by the fact that $H(M_1)=B_1$; and $(f)$ follows from
\begin{align*}
1+J_{d-1}T_{d,1}&=(d+J_{d-1})T_{d,1}=J_{d}T_{d,1}.
\end{align*}

Now assume that \eqref{eq:Induc2} holds for some $k\in[1:d-1]$. Substituting the telescoping result \eqref{eq:P2} into \eqref{eq:Induc2}, we have
\begin{align}
\alpha+J_{d-1}\beta
&\geq J_{d}\sum_{j=1}^{k}T_{d,j}B_j+H(W_1,l_{[2:k+1]}|M^{(k)})+\nonumber\\
& \hspace{13pt} \left[J_{d-k}-(d-k)\right]T_{d,k}H(l^{(k)}|M^{(k)})\nonumber\\
&= J_{d}\sum_{j=1}^{k}T_{d,j}B_j+H(W_1,l_{[2:k+1]}|M^{(k)})+\nonumber\\
& \hspace{13pt} J_{d-1-k}T_{d,k}H(l^{(k)}|M^{(k)}).\label{eq:Temp2}
\end{align}
Further substituting \eqref{eq:P1} into \eqref{eq:Temp2}, we have
\begin{align*}
&\alpha+J_{d-1}\beta\\
& \geq J_{d}\sum_{j=1}^{k}T_{d,j}B_j+H(W_1,l_{[2:k+1]}|M^{(k)})+\\
& \hspace{25pt} J_{d-1-k}T_{d,k+1}H(l^{(k+1)}|M^{(k)})\\
& \stackrel{(a)}{=} J_{d}\sum_{j=1}^{k}T_{d,j}B_j+H(W_1,l_{[2:k+1]},M_{k+1}|M^{(k)})+\\
& \hspace{25pt} J_{d-1-k}T_{d,k+1}H(l^{(k+1)},M_{k+1}|M^{(k)})\\
& \stackrel{(b)}{=} J_{d}\sum_{j=1}^{k}T_{d,j}B_j+(1+J_{d-1-k}T_{d,k+1})H(M_{k+1}|M^{(k)})+\\
& \hspace{5pt} H(W_1,l_{[2:k+1]}|M^{(k+1)})+J_{d-1-k}T_{d,k+1}H(l^{(k+1)}|M^{(k+1)})\\
& \stackrel{(c)}{=} J_{d}\sum_{j=1}^{k}T_{d,j}B_j+J_{d}T_{d,k+1}H(M_{k+1}|M^{(k)})+\\
& \hspace{5pt} H(W_1,l_{[2:k+1]}|M^{(k+1)})+J_{d-1-k}T_{d,k+1}H(l^{(k+1)}|M^{(k+1)})\\
& \stackrel{(d)}{=} J_{d}\sum_{j=1}^{k+1}T_{d,j}B_j+H(W_1,l_{[2:k+1]}|M^{(k+1)})+\\
& \hspace{35pt} J_{d-1-k}T_{d,k+1}H(l^{(k+1)}|M^{(k+1)}),
\end{align*}
which completes the induction and hence the proof of \eqref{eq:Induc2}. Here, $(a)$ follows from the fact that $M_{k+1}$ is a function of $W^{(k+1)}$, which is in turn a function of $\{W_1,l_{[2:k+1]}\}$ and further a function of $l^{(k+1)}$; $(b)$ is due to the chain rule for conditional entropy; $(c)$ follows from the fact that 
\begin{align*}
1&+J_{d-1-k}T_{d,k+1}=(T_{d,k+1}^{-1}+J_{d-1-k})T_{d,k+1}\\
&=\left[\sum_{i=1}^{k+1}(d+1-i)+J_{d-1-k}\right]T_{d,k+1}\\
&=\left(\sum_{j=d-k}^{d}j+J_{d-1-k}\right)T_{n,k+1}=J_{d}T_{d,k+1},
\end{align*}
and $(d)$ is due to the facts that all messages are independent and that $H(M_{k+1})=B_{k+1}$. 

Set $k=d$ in \eqref{eq:Induc2}. By the fact that $H(W_1,l_{[2:d]}|M^{(d)})\geq 0$ and $J_0=0$, we have
\begin{align}
\alpha+J_{d-1}\beta &\geq J_{d}\sum_{j=1}^{k}T_{d,j}B_j. \label{eq:Temp3}
\end{align}
Normalizing both sides of \eqref{eq:Temp3} by $\sum_{k=1}^{d}B_k$ completes the proof of the outer bound \eqref{eq:B2}. %\hfill\IEEEQED
\end{IEEEproof}

\section{Proof of the Propositions}
\subsection{Proof of Proposition~\ref{prop:P1}}

We begin with a simple lemma, which is a consequence of Han's inequality \cite{Han-IC78} and the definition of symmetrical codes.

\begin{lemma}[Han's inequality]
\label{lemma:Han}
For any $k\in[1:d-1]$, $j\in[1:k]$, and $\emptyset \neq \tau\subseteq[k+2:d+1]$, symmetrical MLDR codes must satisfy
\begin{align}
\frac{1}{|\tau|}&H(S_{\tau,k+1}|l^{(j-1)},M^{(k)})\geq \frac{1}{d-k}H(l_{k+1}|l^{(j-1)},M^{(k)}).\label{eq:Han}
\end{align}
\end{lemma}

\begin{IEEEproof}
Consider any two nonempty subsets of $[k+2:d+1]$ of the {\em same} cardinalities, which are denoted as $\tau$ and $\tau'$. If $H(S_{\tau,k+1}|l^{(j-1)},M^{(k)})=H(S_{\tau',k+1}|l^{(j-1)},M^{(k)})$ in any symmetrical MLDR code, then the desired inequality \eqref{eq:Han} will follow directly from Han's inequality \cite{Han-IC78}. 
To prove the desired equality, recall symmetrical MLDR codes preserve joint entropy under any storage-node-index permutation. Consider a permutation $\pi$ where only the indices in $[k+2:d+1]$ are permuted, and $\tau$ are mapped to $\tau'$. The set $l^{(j-1)}$ is invariant under this permutation. Thus the joint entropies involved in \eqref{eq:Han} are indeed preserved under this permutation.
\end{IEEEproof}

The following ``exchange" lemma plays an essential role in the proof of Proposition~\ref{prop:P1}.

\begin{lemma}[Exchange lemma]
\label{lemma:exchange}
For any $k\in[1:d-1]$ and $j\in[1:k]$, symmetrical MLDR codes must satisfy
\begin{align}
& \frac{d+1-j}{d-k}H(l^{(k)}|M^{(k)})+H(l^{(j)}|M^{(k)})\nonumber\\
& \hspace{10pt} \geq \frac{d+1-j}{d-k}H(l^{(k+1)}|M^{(k)})+H(l^{(j-1)}|M^{(k)}).\label{eq:exchange}
\end{align}
\end{lemma}

\begin{IEEEproof}
Since $j \leq k$ by the assumption, we have $d+1-j >d-k$. Thus, we may write $d+1-j=i(d-k)+p$ for some integer $i\geq 1$ and $p\in[1:d-k]$. For any $q\in[1:i-1]$, let 
\begin{align*}
\tau_q:=[j+p+&(q-1)(d-k):j+p+q(d-k)-1].
\end{align*} 
Furthermore, let $\tau_0:=[j:j+p-1]$. Then we have $[j:k]=\bigcup_{q=0}^{i-1}\tau_q$. Next, let us show by induction that
for any $q\in[1:i]$
\begin{align}
&qH(l^{(k)}|M^{(k)})+H(l^{(j)}|M^{(k)})\nonumber\\
& \geq qH(l^{(k+1)}|M^{(k)})+H(S_{\bigcup_{r=0}^{i-q}\tau_r,k+1},l^{(j-1)}|M^{(k)}).\label{eq:TT3}
\end{align}

To prove the base case of $q=1$, note that 
\begin{align*}
H(l^{(j)}|M^{(k)}) &= H(l_j,l^{(j-1)}|M^{(k)})\\
&= H(S_{[j+1:d+1],j},l^{(j-1)}|M^{(k)})\\
&\stackrel{(a)}{=} H(S_{[j:k],k+1},S_{[k+2:d+1],k+1},l^{(j-1)}|M^{(k)})\\
&= H(S_{[j:k],k+1},l_{k+1},l^{(j-1)}|M^{(k)}),
\end{align*}
where $(a)$ follows by swapping $j$ with $k+1$ and the fact that $l^{(j-1)}$ is invariant under such a swap. Further note that $S_{[j:k],k+1}$ is a function of $W_{[j:k]}$, which is in turn a function of $l^{(k)}$. It follows that
\begin{align}
&H(l^{(k)}|M^{(k)})+H(l^{(j)}|M^{(k)})\nonumber\\
&= H(l^{(k)},S_{[j:k],k+1}|M^{(k)})+H(S_{[j:k],k+1},l_{k+1},l^{(j-1)}|M^{(k)})\nonumber\\
& \stackrel{(a)}{\geq} H(l^{(k)},S_{[j:k],k+1},l_{k+1}|M^{(k)})\nonumber\\
&\qquad+H(S_{[j:k],k+1},l^{(j-1)}|M^{(k)})\label{eq:TT1}\\
& \stackrel{(b)}{=} H(l^{(k+1)}|M^{(k)})+H(S_{[j:k],k+1},l^{(j-1)}|M^{(k)})\nonumber\\
& = H(l^{(k+1)}|M^{(k)})+H(S_{\bigcup_{r=0}^{i-1}\tau_r,k+1},l^{(j-1)}|M^{(k)}),
\label{eq:TT2}
\end{align}
where $(a)$ follows from the submodularity of entropy, and $(b)$ follows again from the fact that $S_{[j:k],k+1}$ is a function of $l^{(k)}$. This completes the proof of the base case of $q=1$.

Next, assume \eqref{eq:TT3} holds for some $q\in[1:i-1]$, then
\begin{align}
&(q+1)H(l^{(k)}|M^{(k)})+H(l^{(j)}|M^{(k)})\geq qH(l^{(k+1)}|M^{(k)})\nonumber\\
&\quad +H(S_{\bigcup_{r=0}^{i-q}\tau_r,k+1},l^{(j-1)}|M^{(k)})+H(l^{(k)}|M^{(k)}).\label{eq:TT3.5}
\end{align}
Consider a one-to-one swapping between the elements of $\tau_{i-q}$ and $[k+2:d+1]$, and note that $l^{(j-1)}$ is invariant under such swaps. We can write
\begin{align*}
H&(S_{\bigcup_{r=0}^{i-q}\tau_r,k+1},l^{(j-1)}|M^{(k)})\\
&=H(S_{\bigcup_{r=0}^{i-(q+1)}\tau_r,k+1},S_{[k+2:d+1],k+1},l^{(j-1)}|M^{(k)})\\
&=H(S_{\bigcup_{r=0}^{i-(q+1)}\tau_r,k+1},l_{k+1},l^{(j-1)}|M^{(k)}).
\end{align*}
It follows that
\begin{align}
&H(S_{\bigcup_{r=0}^{i-q}\tau_r,k+1},l^{(j-1)}|M^{(k)})+H(l^{(k)}|M^{(k)})\nonumber\\
&=H(S_{\bigcup_{r=0}^{i-(q+1)}\tau_r,k+1},l_{k+1},l^{(j-1)}|M^{(k)})+H(l^{(k)}|M^{(k)})\nonumber\\
&\stackrel{(a)}{=}H(S_{\bigcup_{r=0}^{i-(q+1)}\tau_r,k+1},l_{k+1},l^{(j-1)}|M^{(k)})+\nonumber\\
&\hspace{13pt} H(l^{(k)},S_{\bigcup_{r=0}^{i-(q+1)}\tau_r,k+1}|M^{(k)})\nonumber\\
&\stackrel{(b)}{\geq} H(S_{\bigcup_{r=0}^{i-(q+1)}\tau_r,k+1},l_{k+1},l^{(k)}|M^{(k)})+\nonumber\\
&\hspace{13pt} H(S_{\bigcup_{r=0}^{i-(q+1)}\tau_r,k+1},l^{(j-1)}|M^{(k)})\nonumber\\
&\stackrel{(c)}{=} H(l^{(k+1)}|M^{(k)})+H(S_{\bigcup_{r=0}^{i-(q+1)}\tau_r,k+1},l^{(j-1)}|M^{(k)}),\label{eq:TT6}
\end{align}
where $(a)$ and $(c)$ are due to the fact that $S_{\cup_{r=0}^{i-(q+1)}\tau_r,k+1}$ is a function of $l^{(k)}$, and $(b)$ follows from the submodularity of entropy. Substituting \eqref{eq:TT6} into \eqref{eq:TT3.5} gives
\begin{align*}
&(q+1)H(l^{(k)}|M^{(k)})+H(l^{(j)}|M^{(k)})\\
& \geq qH(l^{(k+1)}|M^{(k)})+\left[H(l^{(k+1)}|M^{(k)})+\right.\\
& \hspace{13pt} \left.H(S_{\bigcup_{r=0}^{i-(q+1)}\tau_r,k+1},l^{(j-1)}|M^{(k)})\right]\\
& = (q+1)H(l^{(k+1)}|M^{(k)})\nonumber\\
&\qquad\qquad+H(S_{\bigcup_{r=0}^{i-(q+1)}\tau_r,k+1},l^{(j-1)}|M^{(k)}),
\end{align*}
which completes the induction and hence the proof of \eqref{eq:TT3}.

Set $q=i$ in \eqref{eq:TT3}. We have
\begin{align}
&iH(l^{(k)}|M^{(k)})+H(l^{(j)}|M^{(k)})\nonumber\\
& \geq iH(l^{(k+1)}|M^{(k)})+H(S_{\tau_0,k+1},l^{(j-1)}|M^{(k)})\nonumber\\
& = iH(l^{(k+1)}|M^{(k)})+H(l^{(j-1)}|M^{(k)})+\nonumber\\
& \hspace{13pt} H(S_{\tau_0,k+1}|l^{(j-1)},M^{(k)})\label{eq:TT7}
\end{align}
where the last equality follows from the chain rule for conditional entropy. Consider a one-to-one swapping between the elements of $\tau_0=[j:j+p-1]$ and $\tau:=[k+2:k+p+1]$, and note $l^{(j-1)}$ is invariant under such swaps. We can write
\begin{align}
H&(S_{\tau_0,k+1}|l^{(j-1)},M^{(k)})=H(S_{\tau,k+1}|l^{(j-1)},M^{(k)})\nonumber\\
& \stackrel{(a)}{\geq} \frac{p}{d-k}H(l_{k+1}|l^{(j-1)},M^{(k)})\nonumber\\
& \stackrel{(b)}{\geq} \frac{p}{d-k}H(l_{k+1}|l^{(k)},M^{(k)})\nonumber\\
& = \frac{p}{d-k}\left[H(l^{(k+1)}|M^{(k)})-H(l^{(k)}|M^{(k)})\right],\label{eq:TT9}
\end{align}
where $(a)$ follows from Lemma~\ref{lemma:Han}, and $(b)$ is because conditioning reduces entropy. Substituting \eqref{eq:TT9} into \eqref{eq:TT7} gives
\begin{align*}
&\left(i+\frac{p}{d-k}\right)H(l^{(k)}|M^{(k)})+H(l^{(j)}|M^{(k)})\\
& \geq \left(i+\frac{p}{d-k}\right)H(l^{(k+1)}|M^{(k)})+H(l^{(j-1)}|M^{(k)}),
\end{align*}
which is equivalent to \eqref{eq:exchange} by noting that 
\begin{align*}
i+\frac{p}{d-k}=\frac{i(d-k)+p}{d-k}=\frac{d+1-j}{d-k}.
\end{align*}
This completes the proof of Lemma~\ref{lemma:exchange}.
\end{IEEEproof}

Proposition~\ref{prop:P1} can now be readily proved from Lemma~\ref{lemma:exchange} as follows. Fix $k\in[1,d-1]$, add the inequalities \eqref{eq:exchange} for $j\in[1:k]$, and cancel the common term $\sum_{j=1}^{k-1}H(l^{(j)}|M^{(k)})$ on both sides. We have
\begin{align*}
&\frac{\sum_{j=1}^{k}(d+1-j)}{d-k}H(l^{(k)}|M^{(k)})+H(l^{(k)}|M^{(k)})\nonumber\\
& \hspace{10pt}\geq \frac{\sum_{j=1}^{k}(d+1-j)}{d-k}H(l^{(k+1)}|M^{(k)})+H(l^{(0)}|M^{(k)}),
\end{align*}
which can be equivalently written as 
\begin{align}
&\frac{\sum_{j=1}^{k+1}(d+1-j)}{d-k}H(l^{(k)}|M^{(k)})\nonumber\\
&\hspace{20pt} \geq \frac{\sum_{j=1}^{k}(d+1-j)}{d-k}H(l^{(k+1)}|M^{(k)}),\label{eq:Temp4}
\end{align}
by the fact that $H(l^{(0)}|M^{(k)})=0$. Multiplying both sides of \eqref{eq:Temp4} by $d-k$ and writing $\sum_{j=1}^{k+1}(d+1-j)$ and $\sum_{j=1}^{k}(d+1-j)$ as $1/T_{d,k+1}$ and $1/T_{d,k}$ respectively complete the proof of Proposition~\ref{prop:P1}.

\subsection{Proof of Proposition~\ref{prop:P2}}

\begin{lemma}[Han's inequality]
\label{lemma:Han2}
Define $l'_r:=\{S_{1,r},l_r\}$ for $r \in [2:d]$. For any $k\in[1:d-1]$, $j\in[1:k]$, and $\emptyset \neq \tau\subseteq[k+2:d+1]$, symmetrical MLDR codes must satisfy
\begin{align}
\frac{1}{|\tau|}H(S_{\tau,k+1}|l'_{[2:j]},M^{(k)})\geq \frac{1}{d-k}H(l_{k+1}|l'_{[2:j]},M^{(k)}).\label{eq:Han2}
\end{align}
\end{lemma}

The proof follows identical steps to those for Lemma~\ref{lemma:Han} and is omitted due to the space constraint.

\begin{lemma}[Exchange lemma]
\label{lemma:exchange2}
For any $k\in[1:d-1]$ and $j\in[1:k]$, symmetrical MLDR codes must satisfy
\begin{align}
& \frac{d+1-j}{d-k}H(W_1,l_{[2:k]}|M^{(k)})+H(l^{(j)}|M^{(k)})\nonumber\\
& \geq \frac{d+1-j}{d-k}H(W_1,l_{[2:k+1]}|M^{(k)})+H(l^{(j-1)}|M^{(k)}).\label{eq:exchange2}
\end{align}
\end{lemma}

\begin{IEEEproof}
Since $j \leq k$ by the assumption, we have $d+1-j >d-k$. Thus, we may write $d+1-j=i(d-k)+p$ for some integer $i\geq 1$ and $p\in[1:d-k]$. For any $q\in[1:i-1]$, let 
\begin{align*}
\tau_q:=[j+p+(q-1)(d-k):j+p+q(d-k)-1].
\end{align*} 
Furthermore, let $\tau_0:=\{1\}\bigcup[j+1:j+p-1]$. Then we have $\{1\}\bigcup[j+1:k]=\bigcup_{q=0}^{i-1}\tau_q$. Next, let us show by induction that
for any $q\in[1:i]$, we have
\begin{align}
&qH(W_1,l_{[2:k]}|M^{(k)})+H(l^{(j)}|M^{(k)})\nonumber\\
& \geq qH(W_1,l_{[2:k+1]}|M^{(k)})+H(S_{\bigcup_{r=0}^{i-q}\tau_r,k+1},l'_{[2:j]}|M^{(k)}).\label{eq:QQ1}
\end{align}

To prove the base case of $q=1$, note that 
\begin{align*}
H&(l^{(j)}|M^{(k)})\stackrel{(a)}{=} H(l'_{[2:j+1]}|M^{(k)})= H(l'_{j+1},l'_{[2:j]}|M^{(k)})\\
&= H(S_{1,j+1},S_{[j+2:d+1],j+1},l'_{[2:j]}|M^{(k)})\\
&\stackrel{(b)}{=} H(S_{1,k+1},S_{[j+1:k],k+1},S_{[k+2:d+1],k+1},l'_{[2:j]}|M^{(k)})\\
&= H(S_{1,k+1},S_{[j+1:k],k+1},l_{k+1},l'_{[2:j]}|M^{(k)}),
\end{align*}
where $(a)$ follows by swapping $r$ with $r+1$ for all $r\in[1:d]$ and $d+1$ with $1$, and $(b)$ follows by swapping $j+1$ with $k+1$. We thus have
\begin{align}
&H(W_1,l_{[2:k]}|M^{(k)})+H(l^{(j)}|M^{(k)})\nonumber\\
&= H(W_1,l_{[2:k]}|M^{(k)})+\nonumber\\
& \hspace{13pt} H(S_{1,k+1},S_{[j+1:k],k+1},l_{k+1},l'_{[2:j]}|M^{(k)})\nonumber\\
& \stackrel{(a)}{=}H(W_1,l'_{[2:k]},S_{1,k+1},S_{[j+1:k],k+1}|M^{(k)})+\nonumber\\
& \hspace{13pt} H(S_{1,k+1},S_{[j+1:k],k+1},l_{k+1},l'_{[2:j]}|M^{(k)})\nonumber\\
& \stackrel{(b)}{\geq} H(W_1,l'_{[2:k]},S_{1,k+1},S_{[j+1:k],k+1},l_{k+1}|M^{(k)})+\nonumber\\
& \hspace{13pt} H(S_{1,k+1},S_{[j+1:k],k+1},l'_{[2:j]}|M^{(k)})\nonumber\\
& \stackrel{(c)}{=} H(W_1,l_{[2:k+1]}|M^{(k)})+\nonumber\\
& \hspace{13pt} H(S_{1,k+1},S_{[j+1:k],k+1},l'_{[2:j]}|M^{(k)})\nonumber\\
& = H(W_1,l_{[2:k+1]}|M^{(k)})+H(S_{\bigcup_{r=0}^{i-1}\tau_r,k+1},l'_{[2:j]}|M^{(k)}),
\label{eq:QQ2}
\end{align}
where $(a)$ and $(c)$ follow from the facts that $\{S_{1,2},\ldots,S_{1,k+1}\}$ is a function of $W_1$ and that $S_{[j+1:k],k+1}$ is a function of $W_{[j+1:k]}$ which is in turn a function of $\{W_1,l_{[2:k]}\}$, and $(b)$ is due to the submodularity of entropy. This completes the proof of the base case of $q=1$.

Assume that \eqref{eq:QQ1} holds for some $q\in[1:i-1]$. We have
\begin{align}
&(q+1)H(W_1,l_{[2:k]}|M^{(k)})+H(l^{(j)}|M^{(k)})\nonumber\\
& \geq qH(W_1,l_{[2:k+1]}|M^{(k)})+H(S_{\bigcup_{r=0}^{i-q}\tau_r,k+1},l'_{[2:j]}|M^{(k)})+\nonumber\\
& \hspace{13pt} H(W_1,l_{[2:k]}|M^{(k)}).\label{eq:QQ3}
\end{align}
Consider a one-to-one swapping between the elements of $\tau_{i-q}$ and $[k+2:d+1]$, and note that $l'_{[2:j]}$ is invariant under such swaps. We have
\begin{align}
&H(S_{\bigcup_{r=0}^{i-q}\tau_r,k+1},l'_{[2:j]}|M^{(k)})+H(W_1,l_{[2:k]}|M^{(k)})\nonumber\\
&=H(S_{\bigcup_{r=0}^{i-(q+1)}\tau_r,k+1},S_{[k+2:n],k+1},l'_{[2:j]}|M^{(k)})\nonumber\\
& \hspace{13pt} +H(W_1,l_{[2:k]}|M^{(k)})\nonumber\\
&=H(S_{\bigcup_{r=0}^{i-(q+1)}\tau_r,k+1},l_{k+1},l'_{[2:j]}|M^{(k)})\nonumber\\
& \hspace{13pt}+ H(W_1,l_{[2:k]}|M^{(k)})\nonumber\\
& \stackrel{(a)}{=}H(S_{\bigcup_{r=0}^{i-(q+1)}\tau_r,k+1},l_{k+1},l'_{[2:j]}|M^{(k)})\nonumber\\
& \hspace{13pt} +H(W_1,l'_{[2:k]},S_{\bigcup_{r=0}^{i-(q+1)}\tau_r,k+1}|M^{(k)})\nonumber\\
& \stackrel{(b)}{\geq} H(W_1,l'_{[2:k]},S_{\bigcup_{r=0}^{i-(q+1)}\tau_r,k+1},S_{[j+1:k],k+1},l_{k+1}|M^{(k)})\nonumber\\
& \hspace{13pt} +H(S_{\bigcup_{r=0}^{i-(q+1)}\tau_r,k+1},l'_{[2:j]}|M^{(k)})\nonumber\\
& \stackrel{(c)}{=} H(W_1,l_{[2:k+1]}|M^{(k)})+H(S_{\bigcup_{r=0}^{i-(q+1)}\tau_r,k+1},l'_{[2:j]}|M^{(k)}),
\label{eq:QQ4}
\end{align}
where $(a)$ and $(c)$ are true because $\{S_{1,2},\ldots,S_{1,k+1}\}$ is a function of $W_1$ and that $S_{\bigcup_{r=0}^{i-(q+1)}\tau_r,k+1}$ is a function of $W_{\bigcup_{r=0}^{i-(q+1)}\tau_r}$ which is in turn a function of $\{W_1,l_{[2:k]}\}$, and $(b)$ is due to the submodularity of entropy. Substituting \eqref{eq:QQ4} into \eqref{eq:QQ3} gives
\begin{align*}
&(q+1)H(W_1,l_{[2:k]}|M^{(k)})+H(l^{(j)}|M^{(k)})\\
& \geq (q+1)H(W_1,l_{[2:k+1]}|M^{(k)})+\\
& \hspace{13pt} H(S_{\bigcup_{r=0}^{i-(q+1)}\tau_r,k+1},l'_{[2:j]}|M^{(k)}),
\end{align*}
which completes the induction and hence the proof of \eqref{eq:QQ1}.

Set $q=i$ in \eqref{eq:QQ1}. We have
\begin{align}
&iH(W_1,l_{[2:k]}|M^{(k)})+H(l^{(j)}|M^{(k)})\nonumber\\
& \geq iH(W_1,l_{[2:k]}|M^{(k+1)})+H(S_{\tau_0,k+1},l'_{[2:j]}|M^{(k)})\nonumber\\
& \stackrel{(a)}{=} iH(W_1,l_{[2:k]}|M^{(k)})+H(l'_{[2:j]}|M^{(k)})+\nonumber\\
& \hspace{13pt}H(S_{\tau_0,k+1}|l'_{[2:j]},M^{(k)})\nonumber\\
& \stackrel{(b)}{=} iH(W_1,l_{[2:k]}|M^{(k)})+H(l^{(j-1)}|M^{(k)})+\nonumber\\
& \hspace{13pt} H(S_{\tau_0,k+1}|l'_{[2:j]},M^{(k)}),\label{eq:QQ5}
\end{align}
where $(a)$ follows from the chain rule for conditional entropy, and $(b)$ follows by swapping $r$ with $r+1$ for $r\in[1:d]$ and $d+1$ with $1$. Consider a one-to-one swapping between the elements of $\tau_0=\{1\}\bigcup[j+1:j+p-1]$ and $\tau:=[k+2:k+p+1]$, and note $l'_{[2:j]}$ is invariant under such swaps. We can write
\begin{align}
&H(S_{\tau_0,k+1}|l'_{[2:j]},M^{(k)})=H(S_{\tau,k+1}|l'_{[2:j]},M^{(k)})\nonumber\\
& \stackrel{(a)}{\geq} \frac{p}{d-k}H(l_{k+1}|l'_{[2:j]},M^{(k)})\nonumber\\
& \stackrel{(b)}{\geq} \frac{p}{d-k}H(l_{k+1}|W_1,l_{[2:k]},M^{(k)})\nonumber\\
& = \frac{p}{d-k}\left[H(W_1,l_{[2:k+1]}|M^{(k)})-H(W_1,l_{[2:k]}|M^{(k)})\right]\label{eq:QQ6},
\end{align}
where $(a)$ follows from Lemma~\ref{lemma:Han2}, $(b)$ is because $l'_{[2:j]}$ is a function of $\{W_1,l_{[2:k]}\}$. Substituting \eqref{eq:QQ6} into \eqref{eq:QQ5} gives
\begin{align*}
&\left(i+\frac{p}{d-k}\right)H(W_1,l_{[2:k]}|M^{(k)})+H(l^{(j)}|M^{(k)})\\
& \geq \left(i+\frac{p}{d-k}\right)H(W_1,l_{[2:k+1]}|M^{(k)})+H(l^{(j-1)}|M^{(k)}),
\end{align*}
which is equivalent to \eqref{eq:exchange2} by noting that 
\begin{align*}
i+\frac{p}{d-k}=\frac{i(d-k)+p}{d-k}=\frac{d+1-j}{d-k}.
\end{align*}
This completes the proof of Lemma~\ref{lemma:exchange2}.
\end{IEEEproof}

Proposition~\ref{prop:P2} can now be readily proved with Lemma~\ref{lemma:exchange2} as follows. Fix $k\in[1,d-1]$, add the inequalities \eqref{eq:exchange2} for $j\in[1:k]$, and cancel the common term $\sum_{j=1}^{k-1}H(l^{(j)}|M^{(k)})$ on both sides. We have
\begin{align*}
&\frac{\sum_{j=1}^{k}(d+1-j)}{d-k}H(W_1,l_{[2:k]}|M^{(k)})+H(l^{(k)}|M^{(k)})\nonumber\\
& \geq \frac{\sum_{j=1}^{k}(d+1-j)}{d-k}H(W_1,l_{[2:k+1]}|M^{(k)})+H(l^{(0)}|M^{(k)}),
\end{align*}
which is equivalent to \eqref{eq:P2} by the fact that $H(l^{(0)}|M^{(k)})=0$. This completes the proof of Proposition~\ref{prop:P2}.

%\section*{Acknowledgment}

\bibliographystyle{ieeetr}

\end{document}